# $^{89}$Y nuclear spin-lattice relaxation in SrY$_2$O$_4$:Ho$^{3+}$ single crystals

A.V. Dooglav*, I.F. Gilmutdinov, V.I. Katkov, A.G. Kiiamov, I.R. Mukhamedshin

Kazan Federal University, Kremlevskaya 18, Kazan 420008, Russia

*E-mail: Alexander.Dooglav@kpfu.ru

Measurements of $^{89}$Y nuclear spin-lattice relaxation in SrY$_2$O$_4$:Ho$^{3+}$ crystals for the magnetic field **B** parallel to the crystal $c$ axis show that $^{89}$Y at $T < 20$ K relaxes due to fluctuating magnetic field created only by Ho1$^{3+}$ transitions between the ground and the first excited Zeeman sublevels of the $^5I_8$ multiplet.

## 1. Introduction

SrR$_2$O$_4$ compounds in which rare-earth ions R$^{3+}$ occupy two non-equivalent crystallographic positions and are grouped in quasi-one dimensional chains (the space group *Pnam*) attract much attention due to a wide variety of unusual magnetic properties in the ground state. In particular, in SrHo$_2$O$_4$ different magnetic ordering of Ho$^{3+}$ magnetic moments in different chains has been found: short-range order in Ho1 chains and long-range order in Ho2 chains [1-3]. However there are only few papers in which spectroscopic properties of a single ion in the lattice of this compound are studied, and often the results of different papers contradict each other. For example, in [4] inelastic neutron scattering and diffraction experiments show that the ground state of both Ho1 and Ho2 sites in the crystal field are electronic singlets separated from the nearest excited singlets by the energies of about 1 meV for Ho1 and less than 0.3 meV for Ho2. But in [3] it is claimed that the ground state of holmium in both positions is the electronic doublet. Recent optical and high-frequency EPR experiments [5] on diluted SrY$_2$O$_4$:Ho$^{3+}$ single crystals have shown that the ground state of holmium in both positions is the electronic singlet, i.e., the Ho$^{3+}$ ion in the SrY$_2$O$_4$:Ho$^{3+}$ crystal is the Van-Vleck ion.

The additional data on the properties of the ions in singlet ground state can be obtained from the Ho$^{3+}$ enhanced NMR, as it has been done, for example, in [6] for HoF$_3$. But to do it the authors had to build the special NMR spectrometer working at the frequency of few GHz [7].

Some information on the properties of the singlet ground state ions can be obtained from NMR studies of the neighboring diamagnetic ions [8]. The aim of the present work was to study the temperature dependence of $^{89}$Y nuclear spin-lattice relaxation rate in SrY$_2$O$_4$:Ho$^{3+}$ single crystals.

## 2. Crystal growth

Single crystals of Sr(Y$_{1-x}$Ho$_x$)$_2$O$_4$ doped with Ho$^{3+}$ ions (x=0.002 and 0.05) were grown by the optical floating zone technique. SrCO$_3$ (Alfa Aesar, 99.99%), Y$_2$O$_3$ (Alfa Aesar, 99.99%), and Ho$_2$O$_3$ (Alfa Aesar, 99.9%) oxides were used as starting materials. The thoroughly ground and mixed stoichiometric composition was sintered at 1050 °C for 30 hours. Intermediate grindings were performed to obtain a homogeneous composition. The synthesized powder was analyzed by X-ray powder diffraction (XRD) on a Bruker D8 advance (Cu, K$\alpha_1$ and K$\alpha_2$) diffractometer and was found to be a phase-pure material without traces of the of unreacted components or other impurities.

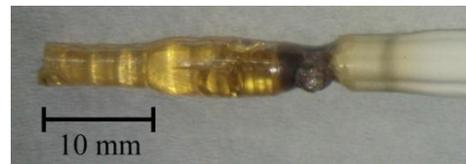

**Figure 1.** SrY$_2$O$_4$ doped with 5% of Ho

The single crystal ingot was grown at a rate of 2 mm/h in an air flow of 0.5 l/min at ambient pressure using an optical floating zone furnace FZ-T-4000-H-VII-VPO-PC (Crystal Systems Corp., Japan) equipped with four 1 kW halogen lamps. Various growth speeds ranging from 1 to 10 mm/h were reported in the previous growth attempts [9]. We have found that growth rate of 2 mm/h led to stable melt zone and allowed to obtain crystal ingots without cracks. The feed and seed rods were rotated at a speed of 15-25 rpm in opposite directions to obtain a homogeneous molten zone. The quality of the crystals were checked with X-rays diffractometry. It appeared that in both crystals the *c* axis was almost perpendicular to the cylindrical boule. One of the as-grown crystals is shown in Fig.1.

## 3. Experimental results

The longitudinal nuclear spin relaxation rate of $^{89}$Y ($I = ½$, $\gamma/2\pi = 0.209$ kHz/Oe, 100% natural abundance) in the temperature range 2.5-180 K was measured by recording the spin echo intensity after the $\pi/2 - \pi$ RF pulses as a function of the repetition time T of this pair of pulses. The obtained $M_Z(T)$ curves were fit to a $M_Z(T)/M_Z(\infty) = 1 - \exp[-(T/T_1)^\beta]$ function with $T_1$ and $\beta$ as fit parameters. The parameter $\beta$ was temperature independent and equal to about 0.7 for 0.2% sample and 0.6 for 5% sample. Since the $T_1$ and $\beta$ values obtained with such measurement method depend on the tip angle (the "true" values are obtained only if the $\pi/2$- and $\pi$-pulses rotate the magnetization $M_Z$ exactly by the angle of $\pi/2$ and $\pi$ [10]), the duration of the $\pi/2$ and $\pi$ pulses were carefully adjusted to a maximum echo intensity. It was additionally checked that the $T_1$ and $\beta$ values obtained by the method described above coincide with those obtained by the much more time consuming standard saturation-recovery method. The measurements were done at the frequency about 18 MHz in the external field about 86 kOe applied along the crystal *c*-axis. The experimental results are shown in Fig.2. At low temperatures the $^{89}$Y nuclear spin-lattice relaxation rate obeys the activation law $T_1^{-1} \sim \exp(-\Delta/T)$ with $\Delta = 31$ K for 0.2% sample and $\Delta = 26$ K for 5% sample. Most likely that slightly different values of $\Delta$ for two samples arise due to small difference of the crystal *c* axes orientation in respect to the external field ***B***.

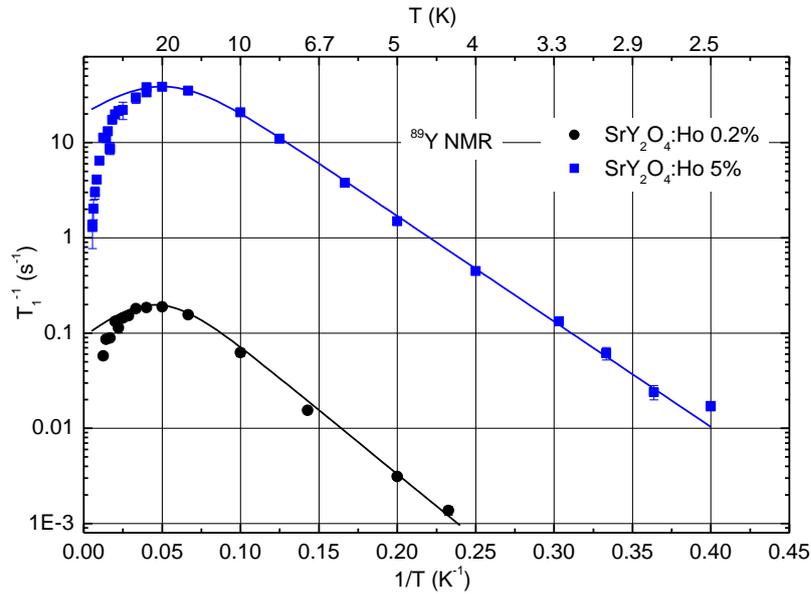

**Figure 2.** Temperature dependence of $^{89}$Y nuclear spin-lattice relaxation rate in $Sr(Y_{1-x}Ho_x)_2O_4$ for two $Ho^{3+}$ concentrations. ***B***||*c*, the resonance frequency $F \approx 18$ MHz, $B \approx 8.6$ T. Solid lines correspond to Eq.(2) with $^{89}\gamma^2 \langle h_1^2 \rangle = 0.45 \cdot 10^8$ s$^{-2}$, $\Delta = 31$ K, $\tau_\infty = 2.16 \cdot 10^{-9}$ s and $^{89}\gamma^2 \langle h_1^2 \rangle = 88 \cdot 10^8$ s$^{-2}$, $\Delta = 26$ K, $\tau_\infty = 2.47 \cdot 10^{-9}$ s for the samples containing 0.2% and 5% Ho, respectively.

## 4. Discussion

The nuclei with diamagnetic electronic shell and nuclear spin $I = 1/2$ ($^{89}Y^{3+}$) actually do not interact with the lattice, and the only mechanism of spin-lattice relaxation for such nuclei in diamagnetic dielectric crystals is their interaction with the lattice via impurity ions with magnetic electronic shell

(via Ho$^{3+}$ ions and other spurious paramagnetic impurities entering the crystal together with holmium in our case). The relaxation transition between the $^{89}$Y Zeeman levels occur due to fluctuating dipolar magnetic field created by magnetic ions at the $^{89}$Y nuclei sites. The magnetic field fluctuates due to changing value and direction of magnetic moments of paramagnetic ions. Since the value of magnetic field created by the magnetic dipole strongly depends on the distance ($H_{dip} \sim r^{-3}$), yttrium nuclei located at a large distance from the magnetic ion (few lattice constants) cannot relax, but are able to transfer their excitation due to nuclear spin diffusion to the nuclei located not far from the magnetic ion (at the diffusion barrier distance [11]).

If the concentration of magnetic ions is large, the diffusion barrier radius becomes greater than the average distance between paramagnetic centers. In this case, all $^{89}$Y nuclei fall into the region of forbidden nuclear spin diffusion. The diffusion barrier radius $b$ is defined as the distance from the paramagnetic ion at which the magnetic field created by its magnetic moment equals to the NMR linewidth $\Delta H_n$ of "distant" $^{89}$Y nuclei [11], i.e., those not affected by the magnetic field of magnetic ion and whose NMR linewidth is determined by nuclear dipole-dipole interaction: $b^3 \sim {}^{Ho}\mu_e/\Delta H_n$. Since $\Delta H_n \sim {}^{Y}\mu_n \cdot N_Y$, and the cube of the average distance $R$ between Ho$^{3+}$ ions $R^3 \sim 1/N_{Ho}$ ($N_Y$ and $N_{Ho}$ being the absolute Y and Ho concentration), $b$ becomes equal to $R$ at $N_{Ho}/N_Y \sim {}^{Y}\mu_n/{}^{Ho}\mu_e$. If to define the diffusion barrier radius as the distance from the paramagnetic ion at which the difference of Ho$^{3+}$ dipole magnetic fields on adjacent yttrium nuclei equals to the NMR linewidth $\Delta H_n$ of "distant" $^{89}$Y nuclei, the marginal holmium concentration becomes larger, of order $N_{Ho}/N_Y \sim ({}^{Y}\mu_n/{}^{Ho}\mu_e)^{3/4}$. Taking ${}^{Y}\mu_n = {}^{89}\gamma \cdot \hbar = 1.39 \cdot 10^{-24}$ erg/G, and the magnetic moment of Ho$^{3+}$ as $\mu_B = 0.927 \cdot 10^{-20}$ erg/G, one gets $N_{Ho}/N_Y \sim 1.5 \cdot 10^{-4}$ in the first case and $\sim 1.4 \cdot 10^{-4}$ in the second. So, the marginal holmium concentration at which all yttrium nuclei fall into the region of forbidden nuclear spin diffusion is of the order 0.14%. Since the reported electronic magnetic moment of Ho$^{3+}$ (in the ordered state) is few Bohr magnetons [4], one can conclude that for even 0.2% Ho sample yttrium nuclear spin diffusion is forbidden, not to mention 5% Ho sample. In this case each yttrium nucleus relaxes by itself via the adjacent paramagnetic centers. The recovery of nuclear longitudinal magnetization after saturation in this case is not exponential, as usual, but obeys the law [12]

$$\frac{M_Z(t)}{M_Z(\infty)} = 1 - \exp\left[-\left(\frac{t}{T_{1n}}\right)^\beta\right] \quad (1)$$

where

$$\frac{1}{T_{1n}} = {}^{89}\gamma^2 \cdot \langle h_1^2 \rangle \cdot \frac{\tau_c}{1+\omega_n^2\tau_c^2}, \quad (2)$$

$\beta = 0.5$, $\omega_n$ – Larmor frequency of $^{89}$Y relaxing nuclei, $\tau_c$ – correlation time of magnetic field fluctuations, $\langle h_1^2 \rangle$ - the mean square of fluctuating magnetic field amplitude. Experimentally obtained $\beta = 0.6$-0.7 is close to the theoretical one.

At very low temperature only the ground state of Ho$^{3+}$ is populated, the magnetic moment of holmium electronic shell is induced by the external magnetic field and does not fluctuate, thus not creating the fluctuating magnetic field at yttrium nuclei. The only channel connecting yttrium nuclei with the lattice in this case is their interaction with "parasitic" paramagnetic impurities. On temperature increasing the excited states of holmium ions become populated. Due to Ho transitions between the ground and excited states the fluctuating field appears with the fluctuations correlation time depending on temperature as $\tau_c = \tau_\infty \cdot \exp(\Delta/T)$, with $\tau_\infty$ the correlation time at infinite temperature, $\Delta$ – the energy of the (nearest) excited state.

At low temperatures the correlation time is large (the fluctuations are slow), $\omega_n \cdot \tau_c \gg 1$, and yttrium nuclear spin-lattice relaxation rate $T_{1n}^{-1} \sim 1/(\omega_n^2 \tau_c) \sim \omega_n^{-2} \cdot \tau_\infty^{-1} \cdot \exp(-\Delta/T)$ is exponentially increasing with temperature. Measuring the rate makes it possible to obtain $\Delta$, the energy of holmium excited state.

At high temperatures the correlation time becomes short, shorter than the Larmor precession period of yttrium nuclear magnetic moment ($\omega_n \cdot \tau_c \ll 1$), and the relaxation rate $T_{1n}^{-1} \sim \tau_c \sim \exp(\Delta/T)$ decreases with temperature increasing and becomes independent on NMR frequency.

At $\omega_n \cdot \tau_c = 1$ the nuclear spin-lattice relaxation rate reaches maximum. Knowing $\Delta$ and temperature at which the rate is maximal, one can obtain $\tau_\infty$. The solid lines in Figure 2 are the plots of the function (2) with parameters shown in the figure caption. At low temperatures the experimental data exactly follow the theoretical curve, while at high temperatures the rate drops more rapidly with temperature than theory predicts. Most likely this discrepancy is caused by the additional, more rapid correlation time decreasing due to populating of $Ho^{3+}$ levels with higher energy, and there maybe by an additional nuclear relaxation channel via parasitic paramagnetic impurities.

Calculations based on optical and ESR experimental data [5] show that for $B \| c$, $B = 8.6$ T the energies of the low lying $Ho^{3+}$ singlet levels are $E = 0$, 32.8, 89.1, 94.4, 98.6 and 142.5 K for Ho1 sites and $E = 0$, 2.5, 52.7, 93.8, 138.5 and 175.2 K for Ho2 sites. Experimentally obtained values of $\Delta = 31$ K for 0.2% sample and $\Delta = 26$ K for 5% sample point to $^{89}Y$ nuclear spin-lattice relaxation at $T < 20$ K due to fluctuating field caused only by Ho1 transitions between the ground and $E = 32.8$ K levels. Why Ho2 transitions between the ground and $E = 2.5$ K levels do not affect $^{89}Y$ relaxation is an open question for us up to now.

## 5. Summary

$^{89}Y$ nuclear spin-lattice relaxation in $SrY_2O_4:Ho^{3+}$ crystals for the magnetic field $\boldsymbol{B}$ parallel to the crystal $c$ axis in the temperature range 2.5 – 180 K was studied. The analysis of the experimental data has shown that $^{89}Y$ at $T < 20$ K relaxes due to fluctuating magnetic fields created only by $Ho1^{3+}$ transitions between the ground and the first excited Zeeman sublevels of the $^5I_8$ multiplet.

**Acknowledgments**

The crystals growth, XRD and NMR measurements were carried out at the Federal Center of Shared Facilities for Physical and Chemical Research of Substances and materials of Kazan Federal University. The authors are grateful to B.Z. Malkin for useful discussions. This work was funded by the subsidy allocated to Kazan Federal University for the state assignment in the sphere of scientific activities (project № 3.8138.2017/8.9).